\documentstyle[epsfig,mathptm,twocolumn]{mn} 
\newif\ifAMStwofonts
\AMStwofontstrue                
\DeclareMathVersion{bold}                    
\DeclareMathAlphabet{\mathbfit}{OT1}{cmr}{bx}{it}
\SetMathAlphabet\mathbfit{bold}{OT1}{cmr}{bx}{it}
\DeclareMathAlphabet{\mathbfss}{OT1}{cmss}{bx}{n}
\SetMathAlphabet\mathbfss{bold}{OT1}{cmss}{bx}{n}
\ifAMStwofonts
  \ifCUPmtlplainloaded \else
    \DeclareSymbolFont{UPM}{U}{eur}{m}{n}
    \SetSymbolFont{UPM}{bold}{U}{eur}{b}{n}
    \DeclareSymbolFont{AMSa}{U}{msa}{m}{n}
    \DeclareMathSymbol{\upi}{0}{UPM}{"19}
    \DeclareMathSymbol{\umu}{0}{UPM}{"16}
    \DeclareMathSymbol{\upartial}{0}{UPM}{"40}
    \DeclareMathSymbol{\leqslant}{3}{AMSa}{"36}
    \DeclareMathSymbol{\geqslant}{3}{AMSa}{"3E}

     \let\le=\leqslant

  \fi
\fi

\newcommand{\apj}{{ ApJ}}

\def\kms{\ifmmode {\rm \ km \ s^{-1}}\else$\rm km s^{-1}$\fi}

\def\gs{\mathrel{\raise0.35ex\hbox{$\scriptstyle >$}\kern-0.6em 
\lower0.40ex\hbox{{$\scriptstyle \sim$}}}}
\def\ls{\mathrel{\raise0.35ex\hbox{$\scriptstyle <$}\kern-0.6em 
\lower0.40ex\hbox{{$\scriptstyle \sim$}}}}

\title
[Investigating the origins of the CMB-XRB cross correlation 
]
{Investigating the origins of the CMB-XRB cross correlation
}
\author[S.P.~Boughn \& R.G.~Crittenden]
{S.P.~Boughn$^1$ \& R.G.~Crittenden$^2$ \\
$^1$Department of Physics, Princeton University, Princeton, NJ 08544\\ and
Department of Astronomy, Haverford College, Haverford, PA  19041 \\
$^2$ Institute of Cosmology and Gravitation, University of Portsmouth,
Portsmouth, PO1 2EG\\
}

\date{\today}
\pagerange{\pageref{firstpage}--\pageref{lastpage}}
\pubyear{2004}
\begin{document}
\maketitle
\label{firstpage}
\begin{abstract} 
Recently, we presented evidence for a cross-correlation of the $WMAP$ satellite map of
the cosmic microwave background (CMB) and the $HEAO1$ satellite map
of the hard X-ray background (XRB) with a dimensionless amplitude
of $0.14 \pm 0.05$ normalized to the product of the $rms$ fluctuations
of the $CMB$ and $XRB$ (Boughn \& Crittenden, 2004).  
Such a correlation is expected in a universe
dominated by a cosmological constant via the integrated Sachs-Wolfe (ISW)
effect and the level of the correlation observed is consistent
with that predicted by the currently favored Lambda cold dark matter
($\Lambda CDM$) model of the universe. 
Since this offers independent confirmation of the cosmological model, it is 
important to verify the origin of the correlation.  Here we explore in detail 
some possible foreground sources of the correlation.  The present evidence all 
supports an ISW origin.   
\end{abstract}


\section{Introduction}
\label{intro}

In a remarkably short time, the standard cosmological model has changed from a
Friedmann universe to a spatially flat universe that is dominated by a cosmological 
constant or some other form of dark energy (e.g., Bahcall et al. 1999).  So far, the primary evidence
for this model comes from supernovae redshift/magnitude observations (Riess et al. 2004) that 
imply the expansion of the universe is accelerating and from the spatial power 
spectrum of the fluctuations in the cosmic microwave background (Bennett et al., 2003).  
While these
pieces of evidence are compelling, especially when combined with the clustering
and density of matter deduced from galaxy observations, it is clearly desirable to
to seek independent confirmation of the $\Lambda CDM$ model.  

The late-time integrated 
Sachs-Wolfe (ISW) effect (Sachs \& Wolfe 1967) promises to provide such an independent test.  
In a $\Lambda$ dominated universe CMB photons 
undergo a net energy shift when traversing linear density perturbations 
(i.e., $\delta \rho / \rho << 1$) at relatively low redshifts ($z \ls 1$). Crittenden and Turok (1996)
suggested that the ISW effect could be detected by correlating the CMB with some
nearby ($z \ls 1$) tracer of matter, e.g., galaxies or AGN. 
This also
occurs in a matter dominated universe with less than the critical density; however,
the effect is spread over larger redshifts (Kamionkowski 1996).  Initial attempts 
to detect such an ISW signal (Boughn, Crittenden, \& Turok 1998; Kamionkowski \&
Kinkhabwala 1999; Boughn \& Crittenden 2002; Boughn, Crittenden \& Koehrsen 2002)
led only to upper limits on $\Lambda$ in a flat $\Lambda CDM$ universe and on the curvature
density in an open, matter dominated universe.  

The situation changed recently
with the release of the first year of data from the $WMAP$ satellite (Bennett
et al. 2003).  Since then, six teams have found correlations of the $WMAP$ CMB data 
with tracers of matter including the hard ($2-10~keV$) X-ray background, the 
NVSS radio galaxy survey, the APM galaxy survey, the Sloan Digital Sky Survey, and 
the 2MASS infrared galaxy survey (Boughn\& Crittenden 2004a; Nolta, et al. 2004; 
Myers et al. 2004; Fosalba \& Gaztanaga 2004; Scranton et al. 2003; Afshordi, Loh,
\& Strauss 2004).  All of these results are only marginally statistically
significant, $\ls 3~\sigma$; however, most of them detect an ISW signal with an 
amplitude similar to that predicted by the $\Lambda CDM$ model.  The X-ray/CMB
cross-correlation (Boughn \& Crittenden 2004) is, perhaps, the most statistically 
significant detection, thanks to the large area covered by the X-ray data and the 
fact the the AGN sources are at somewhat higher redshifts.   
It is the purpose of the present paper to provide the details 
of the X-ray/CMB correlation analysis and to demonstrate that this correlation is 
unlikely to be due to contamination by foregrounds, such as microwave emission from X-ray sources or 
from Galactic emission.

\section{The $HEAO1~A2~2-10~keV$ X-ray Map}

The HEAO1 data set we consider was constructed from the output of two medium energy 
detectors (MED) with different fields of view ($3^\circ \times 3^\circ$ and 
$3^\circ \times 1.5^\circ$) and two high energy detectors (HED3) with these same 
fields of view (Boldt 1987).  The data were collected during the six month period
beginning on day 322 of 1977.  Counts from the four detectors were combined and 
binned in 24,576  $1.3^\circ \times 1.3^\circ$ pixels.  The pixelization we use is 
an equatorial quadrilateralized spherical cube projection on the sky (White and 
Stemwedel 1992).  The combined map has a spectral bandpass (quantum efficiency 
$\gs 50\%$) of approximately $3-17~keV$ (Jahoda \& Mushotzky 1989). For 
consistency with other work, all signals are converted to equivalent flux in the 
$2-10~keV$ band.

Because of the ecliptic longitude scan pattern of the HEAO satellite, sky
coverage and, therefore, photon shot noise are not uniform.  However, the
variance of the cleaned, corrected map, $2.1 \times 10^{-2}~(TOT~counts~s^{-1})^2$,
is significantly larger than the mean variance of photon shot noise, 
$0.8 \times 10^{-2}~(TOT~counts~s^{-1})^2$, 
where $1~TOT~counts~s^{-1} \approx 2.1 \times 10^{-11}
erg~s^{-1} cm^{-2}$ (Allen, Jahoda \& Whitlock 1994).  This implies that most of 
the variance in the X-ray map is due to ``real'' structure.  For this reason and
to reduce contamination from any systematics that might be correlated with the scan
pattern, we chose to weight the pixels equally in this analysis.  However, weighting
the pixels in proportional to their sky coverage makes little difference in the
subsequent analysis.

The resulting map of the hard X-ray background (XRB) has several foreground features
that were fit and removed from the map: a linear time drift of detector
sensitivity, high latitude Galactic emission, the dipole induced by the earth's 
motion with respect to the XRB, and emission from the plane of the local 
supercluster.  These corrections are discussed in detail in Boughn (1999) and in
Boughn, Crittenden, \& Koehrsen (2002).  The map was also aggressively masked by 
removing all pixels within $20^{\circ}$ of the Galactic plane and within 
$30^{\circ}$ of the Galactic center.  In addition, large regions 
($6.5^\circ \times 6.5^\circ$) centered on $92$ nearby, discrete X-ray sources 
with $2 - 10~keV$ fluxes larger than $3 \times 10^{-11} erg~s^{-1} cm^{-2}$ 
(Piccinotti et al. 1982) were removed from the maps.  Around the sixteen brightest of 
these sources (with fluxes larger than $1 \times 10^{-10} erg~s^{-1} cm^{-2}$) 
the masked regions were enlarged to $9^\circ \times 9^\circ$.  

Finally, the map itself 
was searched for ``sources'' that exceeded the nearby background by a specified 
amount.  Since the `quad-cubed' pixelization format lays out the pixels on an approximately square
array, we averaged each pixel with its eight neighbors and then compared this value 
with the median value of the next nearest sixteen pixels (ignoring pixels within the 
masked regions).  If the average flux associated with a given pixel exceeded the 
median flux of the background by a prescribed threshold (1.75 times the mean shot 
noise),  then all 25 pixels ($6.5^\circ \times 6.5^\circ$) were removed from further 
consideration.  The result of all these cuts is a map with $33\%$ sky coverage.  This
was the same map that was used in previous attempts to detect a correlation of the 
the XRB and CMB (Boughn, Crittenden, \& Turok 1998; Boughn, Crittenden, \& Koehrsen
2002) and is used as our canonical X-ray map in the following analysis.  We have
also used other maps with less aggressive source removal and the correlation results
are nearly independent of these cuts; although, the less aggressively masked maps 
result in somewhat higher noise.  For details of the various
source cutting schemes, see Boughn, Crittenden, \& Koehrsen (2002).

\section{The $WMAP$ CMB Map}

The primary CMB map used in the following analysis is the ``internal linear 
combination'' (ILC) map derived from the first year of data from the $WMAP$ satellite 
(Bennett et al. 2003) re-pixelized to be in the same format as the HEAO X-ray map.
The ILC map was constructed so as to have little contamination from the Galaxy; 
however, to avoid any residual low Galactic latitude contamination, we masked it 
with the most aggressive $WMAP$ Galaxy mask ($k0$) from Bennett et al. (2003), 
which results in $68\%$ sky coverage.  For the $1.3^{\circ}$ angular resolution
of the ILC map, instrument noise is negligible so we chose to weight each of the 
unmasked pixels equally in all subsequent analyses.  To check our results we also 
used the independently derived ``cleaned'' map of Tegmark, de Oliveira-Costa, and 
Hamilton (2003).  It is not surprising that this map yielded results that were 
consistent with the ILC map since the two maps were derived from the same primary data.  
To check for a possible frequency dependence of the results we also used Q-band (41 GHz), 
V-band (61 GHz), and W-band (94 GHz) $WMAP$ maps with the same $k0$ masking.  The 
correlation results of 
these three maps are also indistinguishable from the ILC map as will be discussed below.

\section{The CMB/XRB Cross-correlation Function}

\begin{figure}
\vspace{-1.0 cm}
\centerline{\psfig{file=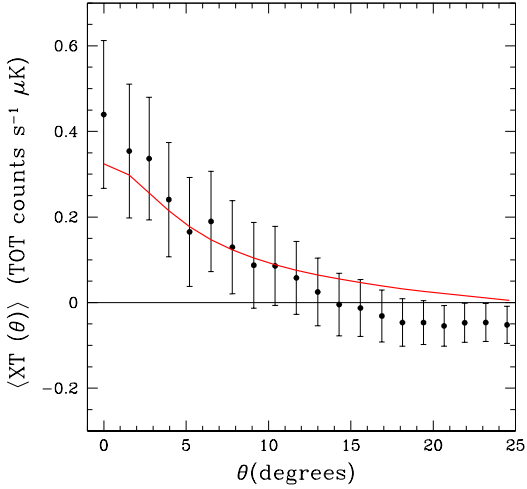,width=3.5in}}
\vspace{-0.5 cm}
\caption{ The points show the cross correlation observed between 
the WMAP and HEAO-A2 data sets.  
The heavy solid line represents the 
predictions for the best fit WMAP $\Lambda CDM$ cosmology.  
} 
\label{fig:ccf} 
\end{figure}

A standard measure of the correlation of two data sets is the cross-correlation
function ($CCF$), which in this case is defined by
\begin{equation}
CCF(\theta) = {1 \over N_{\theta}} \sum_{i,j} (I_i -\bar{I})(T_j -\bar{T})
\label{eq:ccf}
\end{equation}
where the sum is over all pairs of pixels $i,j$ separated by an angle 
$\theta$, $I_i$ is the X-ray intensity of the $i$th pixel, $\bar{I}$ is the mean 
intensity, $T_i$ is the CMB temperature of the $i$th pixel, $\bar{T}$ is the mean 
CMB temperature, and $N_{\theta}$ is the number of pairs of pixels separated by 
$\theta$. The $CCF(\theta)$ generated from the ``cleaned'' map is consistent with
but typically $\sim 10\%$ larger than that generated from the ILC map.  The 
average of these two $CCF$s 
is shown in Figure \ref{fig:ccf}.  The $CCF$ in this figure appears to reveal a 
considerable level of cross-correlation on angular scales $\theta \ls 10^{\circ}$.
However, this is a bit misleading since the error bars in 
Figure \ref{fig:ccf} are highly correlated. 

We estimated the standard deviations and correlation matrix of the errors in the $CCF$ in 
two ways. Using the data themselves, we computed 400 $CCF$s by rotating the ILC and 
``cleaned'' maps with respect to the X-ray map.  
By performing two rotations of angles larger than
$20^{\circ}$, the effects of any intrinsic correlations at small angular scales,
$\theta \ls 10^{\circ}$, are minimized.  Of course, there were pixels
pairs that are repeated in some rotations; however, the number of these are
very small compared to the number of pixel pairs at each angle $\theta$.  While the
signal distribution in the maps is to a good approximation Gaussian, this method
enabled us to estimate the errors in a way that is relatively independent of the 
statistical characteristics of the noise in the maps.  As an alternative we generated 1000 
Monte Carlo CMB maps expected from Gaussian fluctuations with the 
$\Lambda CDM$ cosmological parameters 
consistent with the $WMAP$ data set.  These Monte Carlo maps were cross-correlated with the 
the real X-ray map.  We did not generate Monte Carlo X-ray maps since the XRB is not as
well characterized statistically as the CMB.  That the standard deviations and noise 
correlation matrices of these two methods agree to within a few percent is
an indication that the errors are well characterized to this level.  Since we did not
generate Monte Carlo X-ray maps, there was no correlated component in the Monte Carlo 
trial maps and, therefore, cosmic variance of the ISW signal was ignored.  
Because of the large sky coverage,
the cosmic variance noise of the ISW effect is considerably smaller than other noise
sources and, in any case, does not change the statistical significance of the detection
of a correlated signal in the two maps.

The $CCF$ due to the ISW effect should be achromatic, i.e., independent of the 
frequency at which the CMB is observed (unlike the Sunyaev-Zel'dovich effect).
To check this the X-ray map was cross-correlated with three $WMAP$ CMB maps at
frequencies $41 \pm 4~GHz$ (Q-map), $61 \pm 7~GHz$ (V-map), and $94 \pm 10 GHz$ (W-map).
These three maps were corrected for Galactic emission using the synchrotron, free-free,
and dust maps from the WMAP public data set.  The resulting CCFs are plotted in
Figure 2.  If the positive correlation reported here were due to radio source contamination,
one would expect the Q-band CCF to be considerably larger than the the W-band CCF, which
is clearly not the case.  The case against radio source contamination is discussed in detail
in \S 6.3 below.

\begin{figure}
\vspace{-1.0 cm}
\centerline{\psfig{file=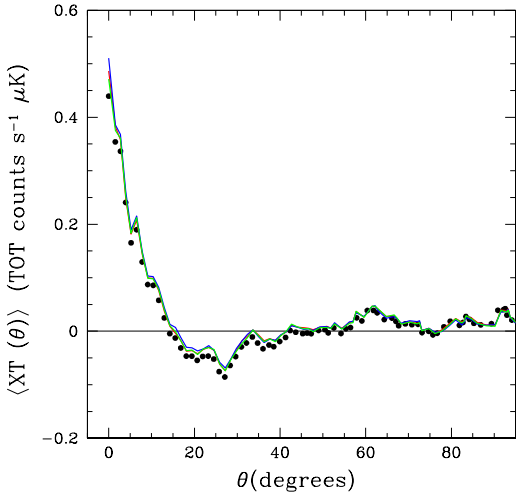,width=3.5in}}
\vspace{-0.5 cm}
\caption{The cross correlations using the various WMAP frequencies 
($Q, V$ and $W$) are all consistent with that found using the 
ILC map.  The nearly coincident red, green and blue curves are CCFs of the X-ray map with
the 41, 61, and 94 $GHz$ CMB maps while the points are from Figure 1.
Much stronger frequency dependence would be expected were the signal 
arising with a typical spectral index expected for radio sources.  
}  
\label{fig:freq} 
\end{figure}

\section{The significance of the detection} 

The first three points 
plotted in Figure \ref{fig:ccf} ($0^{\circ} \le \theta \le 2.8^{\circ}$) are more than $2~\sigma$ 
greater than zero while the next four points ($4.0^{\circ} \le \theta \le 7.8^{\circ}$)
all exceed zero by more than $1~\sigma$.  However, due to the highly correlated nature of the
noise, the overall statistical significance of the detection is only $\sim 3~\sigma$
as will be discussed below.  

Since each of the values of the $CCF$ is determined from a great many pixel pairs (from
$7.2 \times 10^3/$ pairs for the $\theta = 0^{\circ}$ point to $1.5 \times 10^6/$ for 
$\theta = 90^{\circ}$), it is reasonable to assume that the noise in the $CCF$ is Gaussian 
distributed (by the central limit theorem).  Indeed, the observed distributions of the
both the 1000 Monte Carlo trials and the 400 rotated map trials were consistent with Gaussianity.  
As a test of the tails of the distribution we
noted the number of times the $CCF(0)$ values of the noise trials exceeded that level
observed in Figure \ref{fig:ccf}.  Out of the 1000 Monte Carlo trials, the noise exceeded this level 8
times, i.e., $0.008\%$ of the time. This corresponds to a $2.4~\sigma$ effect 
which is consistent with the value of $2.5~\sigma$ of Figure \ref{fig:ccf}.  Similar agreement was 
found for the rotated map trials and for other separation angles.

To determine the overall statistical significance of the detection of the ISW effect 
requires an ISW model.  However, the likelihood that the 
observed $CCF$ is due to noise alone (with no intrinsic correlation of the two maps)
can be evaluated in the absence of any model.  $\chi^2$ provides such an estimate,
i.e.,
\begin{equation}
\chi^2 = \sum_{ij} CCF(\theta_i) (C_{ij})^{-1} CCF(\theta_j)
\end{equation}
where $(C_{ij})^{-1}$ is the inverse of the noise correlation matrix.
$\chi^2$ is a conservative estimate 
of significance since it only assumes Gaussian noise and nothing about the
nature of the signal (other than our expectation 
that the signal occurs at small angles $\theta \ls 10^{\circ}$).
In any case, depending on the number of data points included in the sum,
the null hypothesis is excluded at the $93\%$ to $99\%$ C.L., the latter
of which is for the first 8 data points ($\theta \le  9^{\circ}$) with 
$\chi_{\nu}^2 = 19.5/8$.

The ISW effect provides a prediction for the shape and amplitude of the correlation
for a given $\Lambda$.  Since the shape varies little as $\Lambda$ varies, we can 
use it as a template and fit for the amplitude.  Doing this, the best fit amplitude
of the dimensionless correlation function (normalized to the rms fluctuations of 
the CMB and XRB)
at $\theta = 0^{\circ}$ ranges from $0.118 \pm 0.049$ 
($2.40~\sigma$, 99.2 \% C. L., $\chi^2 = 1.0/1$) for a fit to the first two data points to 
$0.141 \pm 0.048$ ($2.94~\sigma$, $99.8$ C.L., $\chi^2 = 10.9/7$) for a fit to the 
first 8 data points.  We take the latter as our canonical fit and note that the fit to
the first 7 data points gives the same result.  Fits using from 4 to 8 data points 
vary by only $\pm 6\%$.

In the above analysis we have assumed that the noise in both the X-ray and 
CMB maps is Gaussian and this is supported by the pixel 
distribution functions of the X-ray intensity and CMB temperature.  As will be seen below,
the distribution of the product of signals in pixel pairs is also well fit by that 
expected for Gaussian processes as shown in Figures \ref{fig:pdf}  and \ref{fig:resid}a.  
Finally, the 
frequency with which the $CCF(\theta)$ generated by Monte Carlo trials 
exceed a given level are consistent with that expected from Gaussian processes 
out to $3~\sigma$.  The fact that the trials generated from rotated versions of the 
maps agreed with the Monte Carlo trials also indicates that the noise
is statistically well understood.  So from a statistical point of view, our 
$3~\sigma$ results is well characterized by a $0.999\%$ confidence level.

In the present case, the dominant source of
``noise'' is the fluctuations in the CMB from the surface of last scattering and cannot
be easily eliminated.  Even if instrument noise were negligible and 
the Poisson noise associated with the discreteness of the X-ray sources could be
eliminated, one would expect only a factor of 2 improvement in signal to noise
(Crittenden \& Turok 1996). 
It will be some time before the next generation of surveys (e.g., the Large Synoptic
Survey Telescope and the Square Kilometer Array) provides us with new, large-scale
mass tracers that will enable this improvement to be realized.  

\section{Possible Systematic Errors}

Since the significance of the detection is limited, it is essential 
that we carefully 
consider possible systematics which could contribute to the cross correlation. 

\subsection{Localized contamination} 

One important issue is discovering if the signal we observe is 
due to strong correlations between a few contaminated pixels, or is due to 
relatively weak correlations on the whole sky, as predicted by the ISW effect. 
One test of this is to split the data into two hemispheres. The fits of the 
amplitude of the dimensionless correlation function to the north and south Galactic
hemispheres are $0.17 \pm 0.07$ ($\chi^2 = 5.3/7$) and $0.11 \pm 0.07$ ($\chi^2 = 13.8/7$)
using the first 8 data points.  While the noise in the 
detection increases, the signals seen in both hemispheres are clearly consistent with each other. 

\begin{figure}
\vspace{-1.0 cm}
\centerline{\psfig{file=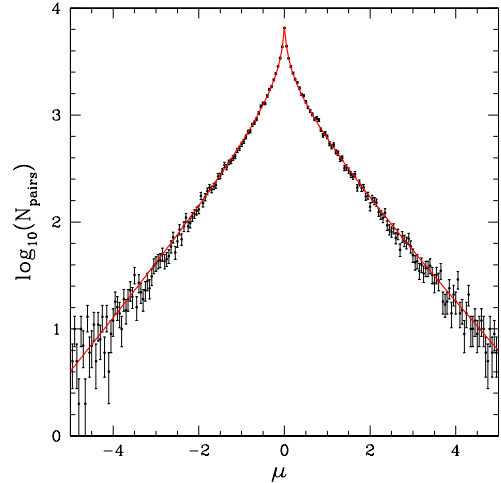,width=3.5in}}
\vspace{-0.5 cm}
\caption{Each bin of the observed correlation in Figure 1 is the average of products of 
pairs of pixels of the X-ray and CMB maps.  This shows the distribution of the 
products for all the pixel pairs contributing to the $2.7^{\circ}$ bin, normalized 
by the {\it rms} values of the two maps.  This figure and the next demonstrate that the 
observed correlation is not arising from a few pixels pairs, but is coming from an 
asymmetry in the full distribution, precisely as expected from a weak correlation of 
two Gaussian fields. 
} 
\label{fig:pdf} 
\end{figure}

\begin{figure}
\vspace{-1.0 cm}
\centerline{\psfig{file=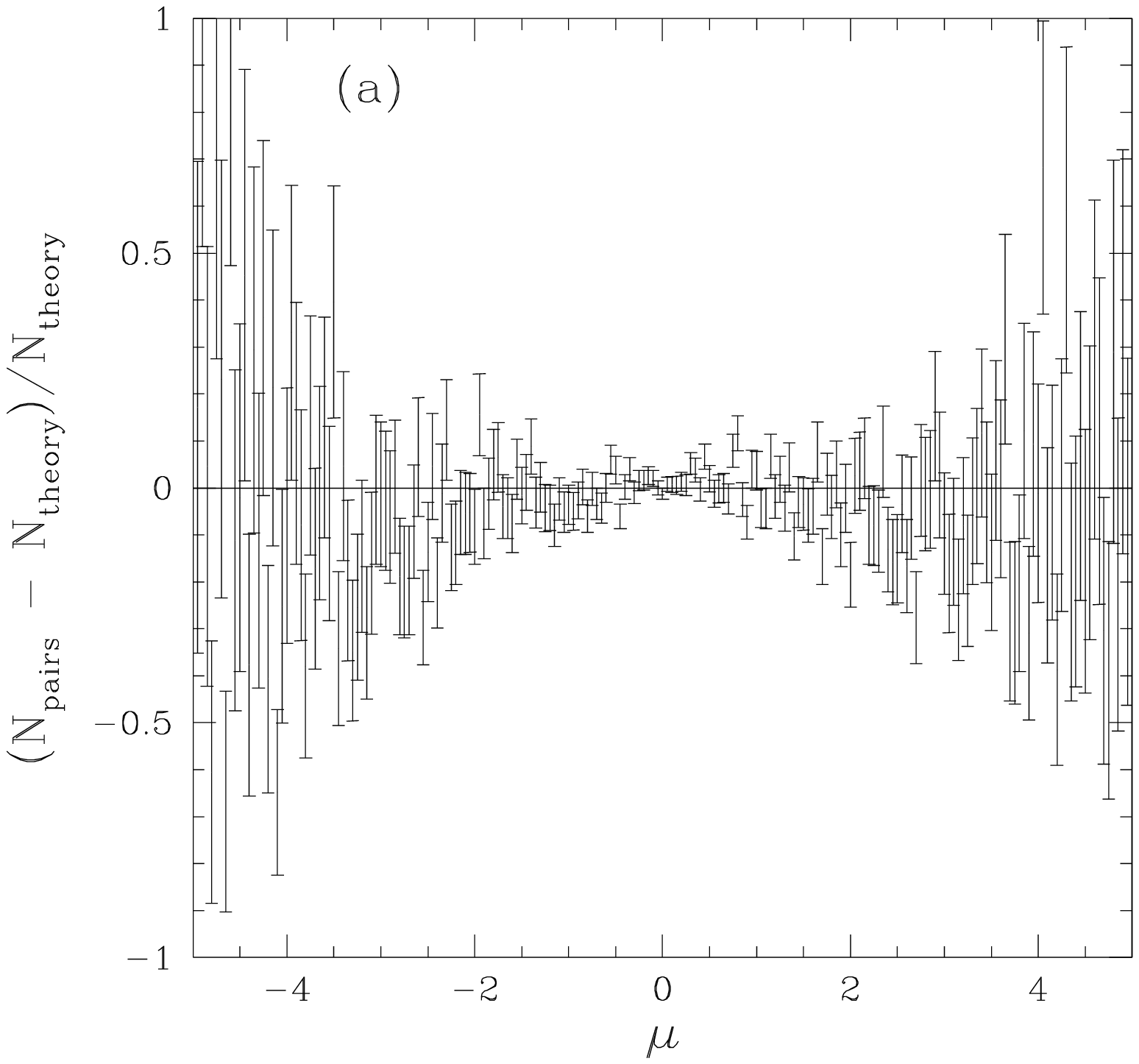,width=3.5in}}
\vspace{-1.5 cm}
\centerline{\psfig{file=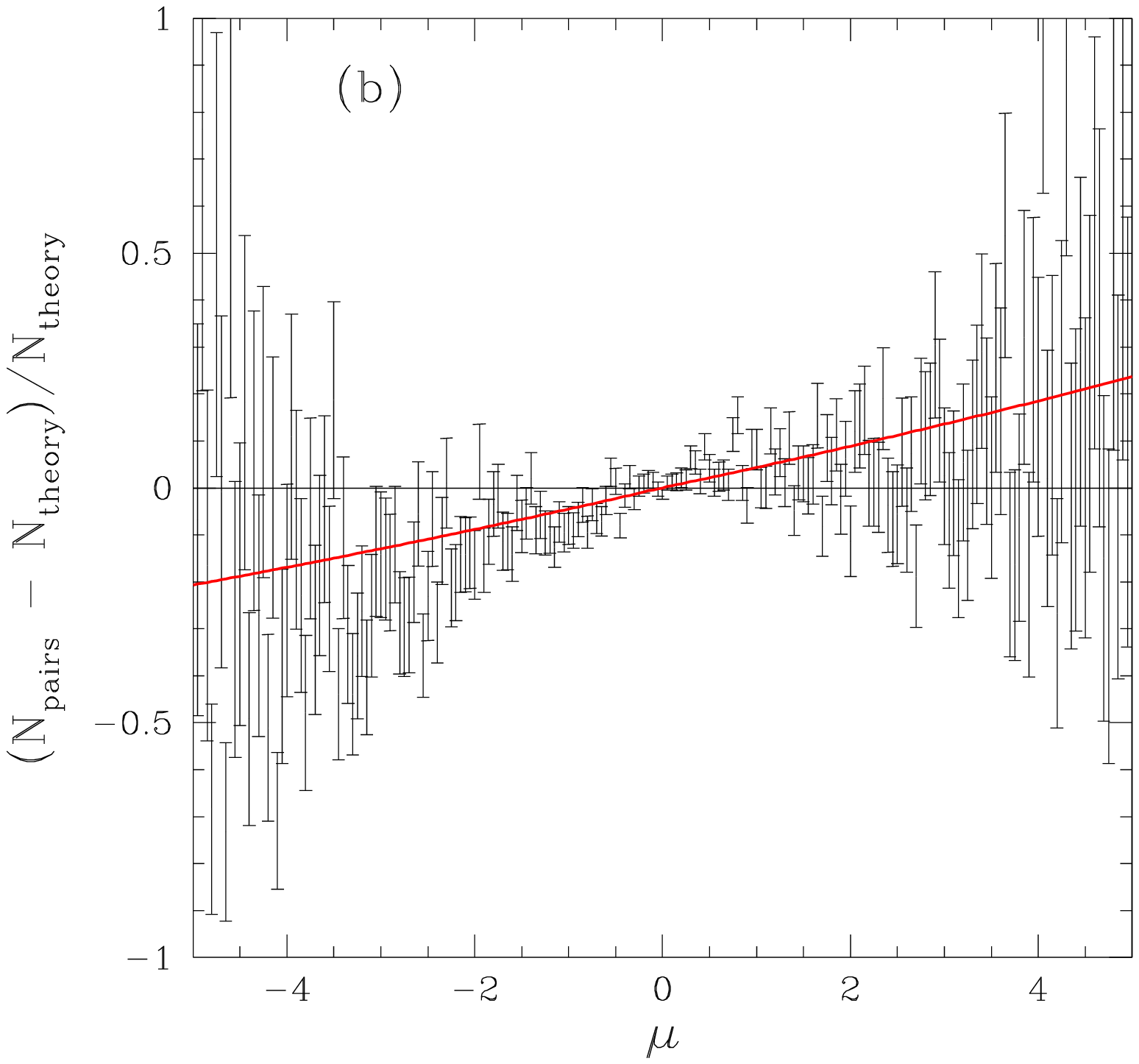,width=3.5in}}
\vspace{-0.5 cm}
\caption{(a) The fractional differences between the observed distribution 
seen in Figure \ref{fig:pdf} and that predicted for the correlation given in Figure \ref{fig:ccf}
at $\theta = 2.7^{\circ}$.  The Poisson error bars are largest at large
$\mu$ where there are the fewest pixels pairs.  (b) The fractional differences
between the observations and the prediction for uncorrelated Gaussian fields.  
The heavy line corresponds to the correlated model used in (a)  
and the data are clearly consistent with the
theoretical curve over the whole range of $\mu$'s.
}
\label{fig:resid}
\end{figure}

Another test for localized contamination is to examine the 
distribution function of the individual terms contributing to the 
$CCF$ of Equation \ref{eq:ccf}, i.e., $(I_i -\bar{I})(T_j -\bar{T})$.  To a 
good approximation, the individual pixel distribution functions of both maps are Gaussian.
If the correlated component of the two maps arises from 
the linear ISW effect, then 
the values of $(I_i -\bar{I})$ and $(T_j -\bar{T})$ should be consistent with 
a bivariate Gaussian distribution with dimensionless cross-correlation coefficient 
$c(\theta) = CCF(\theta)/(\sigma_T \sigma_X)$,  where 
$\sigma_X^2 = \langle (I_i -\bar{I})^2 \rangle$ is the variance of the X-ray intensity
and $\sigma_T^2 = \langle (T_i -\bar{T})^2 \rangle$ is the variance of the CMB
temperature. 
($\sigma_X = 0.124~TOT~cnts~s^-1$ and $\sigma_T = 64 \mu K$ for the maps we used.)   
It is straightforward to show (see Appendix A) 
that the distribution function, $dN/d\mu$, for the product  
$\mu = (I_i -\bar{I})(T_j -\bar{T})/\sigma_T \sigma_X$ is 
given by 
\begin{equation} 
dN/d\mu = {N_{\theta} e^{c(\theta)\mu^{\prime}} K_0(\mu^{\prime}) \over
\pi (1 - {c(\theta)}^2)^{1/2}}
\label{eq:pdf}
\end{equation} 
where $K_0$ is the modified Bessel function, $\mu^{\prime} = \mu/(1-{c(\theta)}^2),$
and $N_{\theta}$ is the number of pixel pairs separated by an angle $\theta$.
Figure \ref{fig:pdf} is a typical plot of the distribution of pixel pairs (in this case for the 
$\theta = 2.7^{\circ}$ bin), as a function of $\mu$ along with the 
theoretical curve given in Equation \ref{eq:pdf}.  The error bars are the usual Poisson error bars, i.e., 
$\sqrt{n}$ where $n$ is the number of pixel pairs in the relevant bin.  
The theoretical curve is not a fit to the distribution function data 
but rather uses the average 
value of $c(\theta)$ determined by Equation \ref{eq:ccf}.  The difference between the observed 
distribution $(dN/d\mu)_{obs}$ and the theoretical distribution of Equation \ref{eq:pdf}
$(dN/d\mu)_{theo}$ for the data of Figure \ref{fig:pdf} is plotted in Figure \ref{fig:resid}a as fractional 
residuals, i.e., $((dN/d\mu)_{obs} - (dN/d\mu)_{theo})/(dN/d\mu)_{theo}$.  
For comparison,
Figure \ref{fig:resid}b is a plot of the fractional residuals assuming $c(\theta) = 0$, i.e.,
in the absence of correlations.  The nearly straight line in Figure \ref{fig:resid}b is the 
theoretical curve with correlations and its slope is a measure of the positive
cross-correlation, $c(\theta)$.  It is clear from this plot that the 
cross-correlation is manifested for the full range of $\mu$.  Were the correlation due to a limited
number of non-Gaussian pixel pairs with large values of $\mu$, one would expect significant 
deviations from the linear behavior exhibited in Figure \ref{fig:resid}b.

\begin{figure}
\vspace{-1.0 cm}
\centerline{\psfig{file=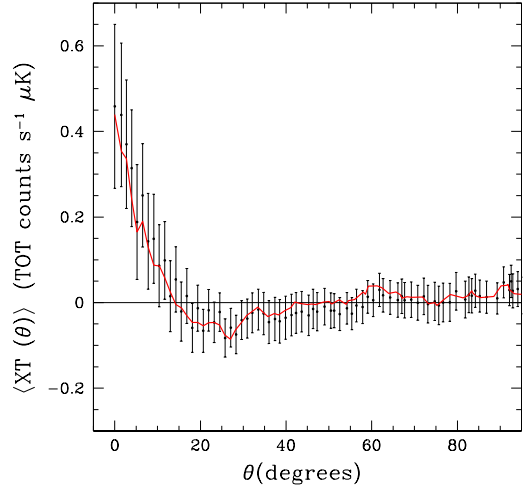,width=3.5in}}
\vspace{-0.5 cm}
\caption{The cross-correlation function estimated by fitting Equation 3 to the distribution function
of pixel pairs (e.g., Figure 3) for all angular separations.  These fits were restricted to 
pixel pair amplitudes of $|\mu| \le 4$.  The results are consistent with the CCF of Figure 1 
(solid line) and imply that the correlation is not arising from a few contaminated pixels.  
} 
\label{fig:corr2} 
\end{figure}

Equation \ref{eq:ccf} determined the $CCF(\theta)$ from a straight average of the 
correlation arising from the pixel pairs.  Alternatively, we can fit the distributions 
(such as Figure \ref{fig:pdf}) for the best dimensionless correlation $c(\theta)$ 
for each angular bin.  To do this, we consider 
a statistic $\sum_i((dN/d\mu)_{obs} (\mu_i) - (dN/d\mu)_{theo} (\mu_i))^2/n_i$
where the sum is over the the bins at $\mu_i$ and $n_i$ is the number of pixel 
pairs in the $i^{th}$ bin, i.e., the Poisson variance.  Minimizing this statistic 
with respect to $c(\theta)$ yields another unbiased
estimate for the cross-correlation function $CCF(\theta)$.  
Figure \ref{fig:corr2} is the average correlation function of
such fits to the ILC and ``cleaned'' CMB maps for a fit range of $-4.0 < \mu < 4.0$.
The $CCF(\theta)$ determined in this way is consistent with 
the $CCF$ derived from Equation \ref{eq:ccf} and depicted in Figure 1.  The errors in Figure 4
were determined from the same 1000 Monte Carlo trials discussed above.
Importantly, these estimates were
insensitive to the domains of the fits; i.e., fitting the function over the
interval $-5.0 < \mu < 5.0$ was consistent with that for the interval $-1.0 < \mu < 1.0$.  
This is evident for the example given in Figure \ref{fig:resid}b where the slope provides a  
reasonable fit to the data over the entire range of $\mu$.  

\subsection{Instrumental systematics} 

In general, cross-correlations are relatively insensitive to unknown systematic 
instrumental errors, especially
if the two data sets are as disparate as the $HEAO$ X-ray map and the $WMAP$ CMB map
considered here.  It is difficult to imagine how systematic errors related to the 
instruments could be correlated with each other.  To be sure, if the systematics are 
extremely large, then even a small correlation between them could generate a significant
$CCF$.  However, any residual systematics in the $WMAP$ data appear to be far below the
level of inherent fluctuations in the CMB (Bennett et al. 2003).  The $HEAO$ X-ray map
does suffer from a significant linear drift in detector sensitivity (Jahoda 1993) but we
have fit for and removed this effect down to a level below that of the intrinsic 
fluctuations of the XRB (Boughn 1998).  Also, the instrument drift results in a signal
proportional to ecliptic longitude and, therefore, represents a larger angular 
scale structure than that in Figure \ref{fig:ccf} .  Thus we conclude it is quite unlikely that 
instrument related systematics are responsible for the observed correlations.

\subsection{Foreground radio sources}
 
More problematic might be the microwave emission of foreground radio sources 
that also emit in hard X-rays.  Certainly radio
sources are highly correlated with the X-ray background (Boughn 1998) and if these
sources also emit significantly at microwave frequencies then a positive correlation
between the CMB and the XRB would be expected.  If this were the case, the $CCF$
would have the same profile as the X-ray auto-correlation function ($ACF$).  However,
the $CCF$ profile in Figure 1 is substantially broader than the X-ray $ACF$
with a half maximum width of $\sim 4^{\circ}$ for the former and $\sim 2^{\circ}$
for the latter (Boughn, Crittenden, \& Koehrsen 2002).  
In addition, the analysis of the distribution function discussed above effectively
eliminates the possibility that the $CCF$ signal is due to a few ``ringers'', i.e.,
a few regions of the sky that are coincidentally bright (or dim) in both the X-ray and
CMB. 

We can attempt to 
directly estimate the microwave emission of the sources of the hard X-ray background. 
However, this is difficult since recent surveys have not been extensive and are limited to 
relatively low frequencies ($< 5$ GHz);  in addition, radio emission is only very roughly 
proportional to X-ray luminosity.  Never-the-less, we have made rough estimates by 
transforming the low frequency radio data to WMAP frequencies assuming a power
spectral index of $\alpha \sim -0.4$, which is the average observed for X-ray
selected radio sources (e.g., Reich et al. 2000).  Using the ROSAT/FIRST analysis
of X-ray bright AGN by Brinkmann et al. (2000), we obtain a rough estimate of
the ratio of the 41 GHz radio flux density to 2-10 keV X-ray flux for these sources, 
$\log(F_\nu / F_X) = -14.2$.  Combining this with the $rms$ fluctuation of the
X-ray background, $1.5 \times 10^{-9}~ergs~s^{-1}cm^{-2}sr^{-1}$ (Boughn, Crittenden,
Koehrsen 2002), the implied correlated component of the 41 GHz CMB temperature 
fluctuations is $\delta T \sim 0.02~ \mu K$.  Such a correlated component would
result in a dimensionless correlation amplitude of $6 \times 10^{-4}$, a factor
of 240 smaller than our observed CCF.  At the other extreme, from the $Chandra$
analysis of Bauer et al. (2002), we deduce a flux ratio of $\log(F_\nu / F_X) = -12.0$.
This implies $rms$ temperature fluctuations of $\delta T \sim 3 ~ \mu K$ and a
dimensionless correlation amplitude of $0.09$, only slightly lower than the observed.
These two estimates bracket those estimated from several other observations of the
radio/X-ray relation (Ciliegi et al. 2003; Georgakakis et al. 2003; Reich et al. 2000).
Furthermore, the implied values at 94 GHz are lower by a factor of 
$(41/94)^{2.4} \sim 0.14$.  Even though these estimates indicate that microwave emission
from sources is not a dominant problem, the data is not yet good enough to make a 
strong claim.

A much stronger case against the possibility of radio source contamination comes from the
achromatic nature of the ISW effect.  The mean radio spectral index of X-ray selected 
sources is $\sim -0.4$ (Reich et al. 2000) while the blackbody spectral index of the CMB 
is $\alpha \simeq +2.0$ in the Rayleigh-Jeans part of the spectrum.  If the CCF we observe 
were due to radio source contamination then one would expect the CCF with the $41~GHz$ 
WMAP map to be $(41/94)^{-2.4} \sim 7$ times larger than the CCF with the $94~GHz$ map.
Even inverted spectrum sources with spectral indices as large as $\alpha \simeq +1.2$ 
would imply a factor of two difference between these two CCFs.  The solid and 
two dashed curves in Figure \ref{fig:freq} are CCFs of the X-ray map with
the 41, 61, and 94 $GHz$ CMB maps while the points are from Figure 1.  It 
is clear that the difference between them is a few percent at most and we therefore
conclude that it is extremely unlikely that the observed correlation is due to radio source 
contamination.

We can also consider how the observed correlations depend on the level of point source 
cuts from the X-ray or CMB maps.  Were bright point sources a dominant contribution, the 
cross correlation should fall off as the cuts remove more sources.  
In our canonical X-ray map, we avoided potential contamination of nearby point sources by
masking pixels with excessive X-ray emission (see above). The resulting masked map
had $33\%$ sky coverage.  We also considered less stringent cuts leaving more sky coverage
(Boughn, Crittenden, \& Koehrsen 2002), 
and the observed cross correlations were largely insensitive to the level of these cuts. 

As an alternative procedure to remove point sources, we also used the
ROSAT All-Sky Survey (RASS) Bright Source Catalog (Voges et al. 1996)
to identify relatively bright sources.  While the RASS survey has somewhat
less than full sky coverage ($92\%$), it has a relatively low flux limit that
corresponds to a $2-10~keV$ flux of $\sim 2 \times 10^{-13}~erg~s^{-1}~cm^{-2}$
for a photon spectral index of $\alpha = -2$.   Every source in the RASS catalog was
assigned a $2-10~keV$ flux from its B-band flux by assuming a spectral
index of $-3< \alpha < -1$ as deduced from its HR2 hardness ratio.  For fainter
sources, the computed value of $\alpha$ is quite uncertain; if it fell outside the
typical range of most X-ray sources, $-3< \alpha < -1$, then $\alpha$ was simply
forced to be $-1$ or $-3$.   It is clear that interpolating RASS flux
to the $2-10~keV$ band is not accurate, so one must consider the level to
which sources are masked with due caution.  However, we are only using these
fluxes to mask bright sources and so this procedure is unlikely to bias the
results.  

We considered maps with ROSAT sources removed at three different
inferred $2-10~keV$ flux thresholds. Thirty-four, high Galactic latitude RASS sources
with fluxes in excess of $3 \times 10^{-11} erg~s^{-1} cm^{-2}$ were identified and
$6.5^\circ \times 6.5^\circ$ regions around each source were masked.  Recall that this flux
level is the nominal level for the Piccinotti sources that are already masked.  The
resulting map had $51\%$ sky coverage.  We also identified and removed sources with
inferred $2-10~keV$ fluxes in excess of $2 \times 10^{-11} erg~s^{-1} cm^{-2}$
($47\%$ sky coverage) and fluxes in excess of $1 \times 10^{-11} erg~s^{-1} cm^{-2}$
($34\%$ sky coverage).  The $CCF$s of these three maps were all consistent with the
$CCF$ in Figure \ref{fig:ccf}.  Even a map with no sources removed other than the bright Piccinotti
sources (Piccinotti et al. 1982) ($56\%$ sky coverage), has a $CCF$ that is also consistent
with that of Figure \ref{fig:ccf}.  Since the observed correlation is insensitive to the flux level
of masked sources, we conclude that it is unlikely that the $CCF$ is contaminated by
radio emission from point X-ray sources.

\subsection{Galactic emission}

It is possible that diffuse microwave/X-ray emission from the Galaxy could be a source of
correlation; however, the $WMAP$ ILC map (Bennett et al. 2003) and ``cleaned'' map 
(Tegmark et al 2003) were both constructed so as to minimize Galactic emission.  In the case 
of the X-ray map, we fit for and removed a small component of high Galactic latitude diffuse 
emission (Boughn 1998).  As is true for instrument drift, any diffuse, high latitude
emission would most likely be on larger angular scales than indicated in Figure \ref{fig:ccf}.  
Never the less, we checked for additional contamination by computing the $CCF$ for a variety 
of Galactic latitude cuts from $\pm 20^\circ$ to $\pm 45^\circ$.  While the noise in the 
latter was larger due to lower sky coverage, the $CCF$s for all the cuts were consistent with 
each other.  Finally, we can again use the achromatic nature of the observed CCF as evidence 
against Galactic contamination.  The CCFs of our canonical X-ray map with the three WMAP
maps (41, 61, and 94 GHz) {\it without} corrections for high latitude Galactic emission also
agree with each other to within a few percent.  This would not be the case if the CCFs were
significantly contaminated by Galactic emission since the spectral index of this emission
$-0.7 > \alpha > -0.3$ is so different from the spectral index of the CMB,
$\alpha = 2.0$.

\section{Interpretation in Terms of the ISW Effect}

To interpret the $CCF$ of Figure 1 in terms of the ISW effect requires a cosmological
model.  We assume a flat, $\Lambda CDM$ universe with the parameters favored by the 
$WMAP$ CMB power spectrum, i.e., $\Omega_m = 0.27$, $\Omega_{\Lambda} = 0.73, 
H_0 = 71~km~s^{-1}Mpc^{-1}$), and $n = 1$, a scale invariant spectrum (Spergel et al. 2003).  
While this model completely determines the ISW effect, the resulting $CCF$ also 
depends on the redshift distribution of the X-ray flux, $dF/dz$, and of the X-ray
bias, $b_x$, defined by
\begin{equation}
b = {\delta \rho_x / \rho_x \over \delta \rho / \rho}
\label{eq:bias}
\end{equation}
where $\rho_x$ is the mean X-ray emissivity, $\rho$ is the mean density
of matter and $\delta$ indicates the $rms$ fluctuations of these densities about
their means.  We use the $dF/dz$ of Boughn and Crittenden (2004b) that was 
generated from the X-ray luminosity function of Steffen et al. (2003)
and Cowie et al. (2003).  The bias
defined in Equation \ref{eq:bias} is typically a function of both
redshift $z$ and of the scale on which the fluctuations are observed.
In the present case, however, the scales associated with the $CCF$ are 
large ($\sim 100~Mpc$) and the redshifts small ($z < 1$).  In this regime 
there are good reasons to believe that the bias is both independent of scale
and relatively small, i.e., $b_x \sim 1$ (Benson et al. 2000; Fry 1996; 
Tegmark \& Peebles 1998). 
The assumptions of 
scale and redshift independence combined with $dF/dz$ completely determine the 
shape of the ISW $CCF$ with the amplitude linearly proportional to $b_x$. In fact,
the shape is relatively insensitive to $dF/dz$ and, therefore, the uncertainty
in the X-ray luminosity function is relatively unimportant. The value of the
X-ray bias is taken to be $b_x = 1.06 \pm 0.16$ as determined from the clustering
of the X-ray background using the same $\Lambda CDM$ model (Boughn \& Crittenden 2004b);
this value is also consistent with recent observations of QSO and AGN clustering
(Croom et al. 2003; Wake et al. 2004). 
The theoretical curve in Figure \ref{fig:ccf} is the expected ISW effect for these parameters, i.e.,
it is not a fit to the observed $CCF$.  The $\chi^2$s of the residuals range from 
0.6/1 for the first datum to 12.4/8 for the first 8 data points.  Beyond $10^\circ$,
the data are consistent with no correlation.  It is clear that the observed $CCF$ is 
consistent with the expected ISW effect.

It is possible to perform a maximum likelihood fit of an ISW model to the data by allowing 
one or more parameters to vary, e.g., $\Lambda$, $dF/dz$, $b_x$, etc.  Currently, $\Lambda$ 
is quite constrained by the WMAP CMB data (Bennett et al. 2003) while, as mentioned above, the shape of
the ISW signal is relatively insensitive to $dF/dz$.  The X-ray bias is, perhaps, the
least well known of the model parameters so we allowed it to vary to minimize $\chi^2$.
The values of bias so obtained ranged from $b_x = 1.31 \pm 0.55$ for a fit to the
first 3 data points to $b_x = 1.58 \pm 0.54$ for a fit to the first 8 data points.  The
significance of the detections ranged from $2.4\sigma$ to $2.9\sigma$, which provide
another measure of the statistical significance of the detection.  These values 
are somewhat higher than but consistent with the $b_x = 1.06 \pm 0.16$ value  
we derived  (Boughn \& Crittenden 2004b) from the X-ray auto-correlation function. 
In any case, the errors are large and this is certainly not 
the best method to determine either $\Lambda$ or $b_x$.  Since the amplitude of the ISW
correlation is proportional to the bias, the above fits provide a useful way of
characterizing the amplitude of the observed $CCF$.  For example, a fit of the bias to
the first 8 data points implies a dimensionless correlation of 
$CCF/{\sigma_x \sigma_T} = 0.14 \pm 0.05$ where $\sigma_x$ and $\sigma_T$ are the
$rms$ fluctuations of the $XRB$ and $CMB$.

To extract cosmological information requires assumptions about the scale and redshift 
dependence of the bias, as well as knowledge of $dF/dz$.  If the bias is constant in 
both scale and redshift, then the predictions for the dimensionless cross correlation 
will be independent of the bias.  However, some uncertainty in the theoretical predictions
still arise from the uncertainty in $dF/dz$; the amplitude varies about 10-20\% 
for different reasonable assumptions for the redshift distribution.  This is to be added to the 
statistical uncertainties of $\sim 35 \%$.  The fact that the observed value is slightly higher than the 
$\Omega_\Lambda= 0.73$ model predictions suggest the data prefer slightly higher $\Lambda$, but not 
dramatically higher.  However, these weak detections still allow the exclusion of more radical models, 
where the correlation is expected to be much larger (if the matter density is very low) or 
models where a negative correlation is expected, such as very closed models (Nolta et al. 2004).

\section{Comparison with Previous Results}

Essentially the same analysis as described above was performed in a comparison of the
$HEAO$ X-ray map with the $COBE$ satellite CMB map (Boughn, Crittenden \& Koehrsen 2002).  The correlated 
signal evident in Figure \ref{fig:ccf} is large enough to have been marginally detected in that analysis
albeit with smaller signal to noise because of that map's significant
instrument noise and lower angular resolution.  However, we saw no such correlation and 
proceeded to set a $95\%$ C.L. upper limit of the ISW effect at a level lower than the 
positive detection claimed in this paper.  To search for the source of this discrepancy we
constructed a combination map consisting of $25\%$ of the Q-band $WMAP$ map, 
$25\%$ of the V-band map, and $50\%$ of the W-band map suitibly convolved with the 
COBE antenna pattern.   This map should be a good approximation to the $COBE$ map used 
in the previous analysis. When the smoothed combination map is put through 
the same ``pipeline'' as the earlier $COBE$ analysis, we detect a statistically significant
$CCF$ that is consistent with the present result.  While 
the different results for these two analyses is a bit surprising it is not totally 
unexpected. A $2~\sigma$ noise fluctuation in the COBE data would account for the difference.

\begin{figure}
\vspace{-1.0 cm}
\centerline{\psfig{file=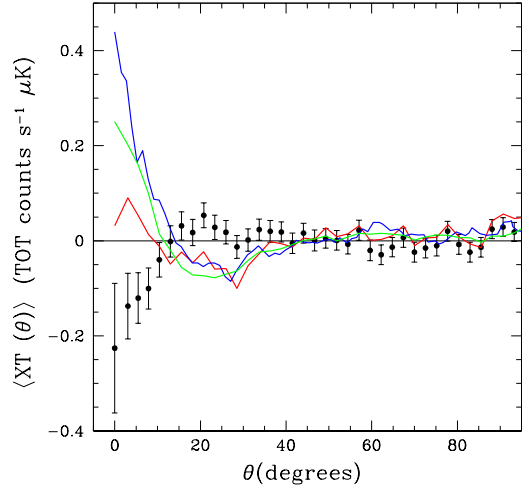,width=3.5in}}
\vspace{-0.5 cm}
\caption{The earlier observations using the four year COBE data
showed much less correlation, apparently because of noise in the maps. 
The correlation seen with WMAP (blue), is suppressed when WMAP is smoothed with the COBE beam (green); 
however, it is still much higher than 
the correlation with observed with COBE (red).  While (COBE - WMAP) should be consistent 
with noise, it is apparently anti-correlated with the X-ray background (points with error bars). 
The reasons 
are unclear, but could indicate a low level systematic in the COBE data.    
}
\label{fig:cobe}
\end{figure}

Furthermore, it is 
quite possible that the $COBE$ map contains some unknown, low amplitude systematic structure.
The $CCF$ of the X-ray map with the difference of the $COBE$ and smoothed combination $WMAP$ 
maps is shown in Figure \ref{fig:cobe}.  It is clear from the plot that there is some systematic 
difference between the two CMB maps on angular scales $\ls 10$ degrees.  Dividing this $CCF$ 
by the $rms$ fluctuations of the X-ray map gives an indication of the correlated differences
between the two CMB maps, i.e., $CCF / \sigma_x \sim 1$ to $2~ \mu K$ compared to the 
$71~\mu K$ instrument noise per pixel in the COBE map.  This level of
discrepancy would have gone unnoticed in previous comparisons of the two maps.  However,
it is not possible to determine whether the discrepancy is due to such a systematic
or is simply ``unlucky'' but statistically possible noise fluctuations in the COBE
map.  In any case,
we are led accept the more statistically significant result with the much cleaner $WMAP$ 
data set presented here.

\section{Conclusions}

We conclude that the integrated Sachs-Wolfe effect has been detected at the $\sim 3~\sigma$ 
level.  There is considerable evidence that the detection is not due to spurious systematics 
or contamination by point radio sources.  
If so, then these and possibly other recent observations 
(Boughn\& Crittenden 2004; Nolta, et al. 2004; Myers et al. 2004; Fosalba \& Gaztanaga 2004; 
Scranton et al. 2003; Afshordi, Loh,\& Strauss 2004 ) offer the first direct glimpse into 
the production of CMB fluctuations and provide important, independent confirmation of the 
new standard cosmological model: an accelerating universe, dominated by dark energy.  It 
should be pointed out that measurements of the power spectrum of CMB fluctuations do not show 
evidence of increased power on large angular scales ($\theta > 20^{\circ}$) as predicted by 
the ISW effect, but rather indicate that there may be power missing on large angular
scales (Spergel et al. 2003).  This deficit is intriguing 
and may be telling us something about the formation of the very largest structures in the 
universe.  The consequences of the ISW effect reported in this letter are primarily
on smaller angular scales and are not in direct conflict with the larger angular scale 
power deficit.

\subsection*{ACKNOWLEDGMENTS}
We thank Ed Groth and Greg Koehrsen for a variety
of analysis programs and Bob Nichol for useful conversations.   
We acknowledge the use of the Legacy Archive for Microwave Background Data Analysis
(LAMBDA). Support for LAMBDA is provided by the NASA Office of Space Science.

\bibliographystyle{/opt/TeX/tex/bib/mn}

\section*{APPENDIX A}

Here we briefly derive Equation \ref{eq:pdf} for the probability distribution of 
the product of two correlated Gaussian distributions.  We begin by assuming that 
we have two Gaussian variables, $x$ and $y$, with unit variance and zero mean, 
and are described by the probability distribution: 
\begin{equation}   
P(x,y)=
\frac{1}{(2\pi)\,\det|\tilde{C}|^{1/2}} \exp \left[ -\frac{1}{2}(x,y)\tilde{C}^{-1}(x,y)^T \right].
\end{equation}   
The dimensionless correlation is given by the variable $c = \langle x y \rangle$, and the 
corresponding covariance matrix is 
\begin{equation}   
\tilde{C} = \left( \begin{array}{cc} 1 & c \\ c & 1 \end{array}\right). 
\end{equation}   

The distribution of the product $z = xy$ is given by the marginalization of the distribution
given the constraint: 
\begin{equation}
P(z = xy) = \int_{-\infty}^{\infty} dx \int_{-\infty}^{\infty} dy \, \delta(xy - z) \, P(x,y). 
\end{equation}
Rewriting the Dirac delta function, $\delta(xy - z) = 1/x \, \delta(y - z/x)$ if  $x > 0$, 
we can perform the 
$y$ integral, leaving: 
\begin{equation}
P(z)  = 2 \int_{0}^{\infty} \frac{dx}{x} P(x, z/x). 
\end{equation}
Substituting the above form for the two-dimensional distribution function, we find
\begin{equation}
P(z)  = \int_{0}^{\infty} \frac{dx}{x} \frac{1}{\pi (1-c^2)^{\frac{1}{2}}} 
\exp \left[ - \frac{1}{2} (x^2 - 2cz + \frac{z^2}{x^2})/(1-c^2) \right] 
\end{equation}
which can be rewritten as
\begin{equation}
P(z)  = \frac{e^{cz/(1-c^2)}}{\pi (1-c^2)^{\frac{1}{2}}}  \int_{0}^{\infty} \frac{dx}{x} 
\exp \left[ - \frac{1}{2} (x^2 + \frac{z^2}{x^2})/(1-c^2) \right].  
\end{equation}
If we make the substitutions $x = e^t$ and $z^2 = e^{2t_0}$, the integral can be written as 
\begin{equation}
P(z)  = \frac{e^{cz/(1-c^2)}}{\pi (1-c^2)^{\frac{1}{2}}}  \int_{-\infty}^{\infty} dt
\exp \left[ - \frac{1}{2} (e^{2t} + e^{2t_0 -2t})/(1-c^2) \right].
\end{equation}
Finally, defining $t' = 2t - t_0$, the final probability distribution can be found: 
\begin{eqnarray}
P(z)  & = & \frac{e^{cz/(1-c^2)}}{\pi (1-c^2)^{\frac{1}{2}}}  \int_{-\infty}^{\infty} \frac{dt'}{2}
\exp \left[ - \cosh(t') |z|/(1-c^2) \right] \\
& = & \frac{e^{cz/(1-c^2)}}{\pi (1-c^2)^{\frac{1}{2}}} K_0 \left( |z|/(1-c^2) \right), 
\end{eqnarray}
where $K_0$ is the modified Bessel function. 
(In evaluating the final integral, we have used Eqn. 3.337, Gradshteyn \& Ryzhik 1980.) 


\label{lastpage}
\bsp 

\end{document}